\documentclass[useAMS,usenatbib]{mn2e}
\usepackage{txfonts}
\usepackage{natbib}
\usepackage{longtable}
\usepackage[dvips]{graphicx}
\usepackage{aas_macros}

\def\Rsolar{$R_{\odot}$}
\def\Msolar{$M_{\odot}$}

\newcommand{\thickhline}{\noalign{\hrule height 0.8pt}}

\title[Mass-transfer variability]{The
influence of mass-transfer variability on the growth of white dwarfs, and the implications for supernova type Ia rates}
\author[]{S. Toonen$^{1,2}$, R. Voss$^{1}$, C. Knigge$^{3}$\\
$^{1}$Department of Astrophysics/IMAPP, Radboud University Nijmegen, PO Box 9010, NL-6500 GL Nijmegen, the Netherlands\\
$^{2}$Leiden Observatory, Leiden University, PO Box 9513, NL-2300 RA Leiden, the Netherlands\\
$^{3}$University of Southampton, School of Physics and Astronomy, Southampton SO17 1BJ, UK\\
}
\begin{document}

\date{}

\pagerange{\pageref{firstpage}--\pageref{lastpage}} \pubyear{2014}

\maketitle

\label{firstpage}

\begin{abstract}
White dwarfs (WDs) can increase their mass by accretion from companion stars,
provided the mass-accretion rate is high enough to avoid nova eruptions. 
The accretion regimes that allow growth of the WDs are usually calculated
assuming constant mass-transfer rates. 
However, it is possible that these systems are influenced by effects that cause
the rate to fluctuate on various timescales.
We investigate how long-term mass-transfer variability affects 
accreting WDs systems.
We show that, if such variability is present, it expands the parameter space of binaries where the WD can effectively increase its mass. 
Furthermore, we find that the supernova type Ia (SNIa) rate is enhanced by a factor 2-2.5 to a rate that is comparable with the lower limit of the observed rates.   
The changes in the delay-time distribution allow for more SNIae in stellar populations with
ages of a few Gyr. 
Thus, mass-transfer variability gives rise to a new formation channel of SNIa events that can significantly contribute to the SNIa rate.
Mass-transfer variability is also likely to affect other binary populations through enhanced WD growth. For example, it may explain why WDs in cataclysmic variables are observed to be more massive than single WDs, on average. 

\end{abstract}


\section{introduction}

White dwarfs (WDs) in binaries can accrete from their companion stars.
Such binaries are called cataclysmic variables (CVs) if the donor stars
are low-mass main-sequence stars, symbiotic binaries (SBs) if they are
evolved red giants, or AM CVNs if the donor stars are low-mass
Helium WDs or Helium stars.
For CVs and SBs, the matter accreted by the WD consists mainly of
hydrogen. As the matter piles up on the surface of the WD, it eventually
reaches temperatures and densities high enough for nuclear burning.

The burning can proceed in two ways, depending on the accretion rate
and the mass of the WD. For high accretion rates and WD masses, the
hydrogen burning on the surface of the WD is continuous 
\citep{Whe73,Nom82}, whereas for
low accretion rates and WD masses the hydrogen is burned in
thermo-nuclear runaway novae \citep{Sch50,Sta74}. 
In general, the high
mass-transfer rates needed for continuous surface hydrogen burning
can only be reached by SBs, where high mass-transfer rates can be
driven by the expansion of the evolved star and by systems with
main-sequence donors more massive than the accreting WDs
\citep{Nom00}.
The masses of 
WDs with high accretion rates
can grow effectively, but at very high accretion rates close to the Eddington limit, the growth of the white dwarf is
limited. 
At these rates a hydrogen red-giant-like envelope forms
around the WD and hydrogen burning on top of the
WD is strong enough for a wind to develop from the WD \citep{Kat94,Hac96}.
On the other hand, at low accretion rates mass accretion on to the WD is not very efficient either, as the nova eruptions eject some or all of the accreted matter from the binary system, possibly along with some of the surface material of the WD itself \citep[e.g.][]{Pri95}.
The average mass-transfer rate allowing growth of the white dwarf is
therefore limited to a relatively narrow range (approximately $10^{-7}-10^{-6}M_{\odot}$
yr$^{-1}$).
 
The growth of WD masses can have important consequences. In the
single-degenerate (SD) scenario for type Ia supernova (SNIa)
progenitors \citep{Whe73,Nom82} 
the accretion on to a carbon-oxygen WD pushes the mass 
above the critical mass limit for WDs (close but not equal to 
the Chandrasekhar limit) which then explodes as a SNIa. In this 
scenario, it is necessary for the WD to retain several tenths of
solar masses of accreted material. It is not possible to achieve
such mass growth for the majority of systems with mass-transfer 
rates in the nova regime, even if some of the accreted matter is retained.
Following this theory, the rate and delay time distribution (DTD, evolution
of the rate as a function of time after a single star formation episode)
can be estimated with the use of population synthesis models 
\citep[e.g.][]{Yun94,Too12,Bou13}.
While there
currently is no consensus between the models as to the shape of the DTD \citep{Nel13, Bou13},
the majority of models agree on two problems: 1) There are not enough
systems with high mass-transfer rates to account for all the observed
SNeIa, and 2) after an age of approximately $6-7$ Gyr, it is not possible to
create SNIa explosions through this scenario, as only low-mass donors
remain.

The considerations above apply to systems where the mass-transfer
rate is given by the evolutionary state of the system. I.e. two
binaries with the same parameters will have the same mass-transfer
rate. Observations of accreting WD systems indicate that the long-term 
average mass-transfer rates do indeed follow the expectations 
\citep[e.g.][]{Kni11}. However, it is possible 
that the mass-transfer
rates are highly variable on intermediate timescales 
\citep{Pat84,Ver84,War87,Ham89}. In this paper,
we discuss such variability and show that it affects 
the evolution of accreting WD systems. In particular, the
effects can be of high importance for understanding SD SNIa
progenitors, as it increases the volume of the parameter space of systems that
can explode as SNeIa.

\section{Mass-transfer variability}
\label{sec_ch3:mt_var}
Over the past decades there have been many studies discussing
the theoretical and observational aspects of mass-transfer
variability in WD binaries. Below we shortly review the current
knowledge in order to construct models that capture the main
effects of the possible variability.
For a thorough review, see \citet{Kni11}, section 4.

\subsection{Theoretical considerations}
In the majority of accreting WD binaries (excepting strongly
magnetic WDs with low accretion rates), hydrogen rich matter
is accreted through an accretion disk that deposits the matter on to the
surface of the white dwarf. The matter quickly spreads over
the surface of the white dwarf. What then happens depends on
the rate of accretion and the resulting temperature and density 
structure near the surface of the white dwarf \citep{Nom82,
Nom07,She07}. 
At low accretion
rates the temperature of the white dwarf surface remains low and
the accreted hydrogen burns in an unstable manner, leading to
nova eruptions that eject most (if not all) of the accreted matter.
At high accretion rates the hydrogen burning is stable and the
matter remains on the WD, except if the accretion rate is so high
(roughly the Eddington limit) that most of the matter cannot
be retained by the WD. 

For the fate of the accreted matter in a system with mass-transfer
variability to be different for a similar system without variability,
the timescale must neither be too long nor too short. If it is
too long, the properties of the binary that depend on the average long-term 
mass-transfer rate are affected, such as the radius of the donor star.  
This would be observable and would also change the
whole evolution of the binary \citep[see e.g.][]{Kni11}. 
On the other hand, if the timescale
of the variability is too short, the surface temperature of the
white dwarf is not adjusted to the instantaneous mass-transfer
rate, which is necessary for the burning to be affected. 

For example, in the accretion disk instability model \citep[e.g.][]{Osa96, Las01}, the mass 
transfer rate is increased by a factor of approximately $10^3-10^5$ during 
outbursts (observed as dwarf novae), and this model has been invoked 
to stabilize the hydrogen burning \citep{Kin03, Ale11}.
However, in this model, the accretion rate is only high for a
very short time, 
and the heat and density of the accreted layer
is not raised enough during the outburst to ignite \citep{Tou05}, as also evidenced by the lack of hydrogen burning
events triggered by dwarf novae. 
Therefore, the layer builds up
without a significant temperature increase, and when burning
eventually is ignited, it is unstable and therefore leads
to a nova eruption\footnote{The effect of the instability of accretion discs on the SNIa rate has been studied by \citet{Wan10} with a model similar to our model CONST. If stable burning can be maintained, the SNIa rate is increased by a factor 2-3 compared to a model without mass-transfer variability. Note that while the effect on the SNIa rate is similar to our findings, the model that we assume (model NORM-MAX) is different. }.

Another example regards nova eruptions. After the eruption, the temperature
of the white dwarf is increased, and it is possible that
for a short time the burning can be stable. Such short-lived stable surface burning triggered by
nova eruptions is seen in some systems \citep[see][and Sect.\,\ref{sec_ch3:mt_var_obs}]{Sch10}, 
but radiation losses during the quiescent periods quickly cool down the WD into the unstable burning regime\footnote{Note that in the context of eq.\,\ref{eq:r_mt}~and~\ref{eq:r_off}, this means that $f\approx 1$ and that the majority of the mass is transferred in the off-state at low mass-transfer rates.}. 

Thus, if mass-transfer variability is to significantly
change the surface burning, the timescale of the mass-transfer
fluctuations must at least be longer than the timescale of
the eruption. In this case a continuous high accretion
rate after the eruption ensures that the temperature on the surface of the WD is sufficient such that the nuclear burning continues.
It also means that only
the variability of the rate of matter
being transferred from the companion star to the WD accretion disk 
can be of importance to the growth of the WD (and not the variability of the transfer from the disk to the WD).
Such a variability can be achieved in two ways, either through the 
change in the radius of the companion star, or through a change in 
the size of its Roche-lobe \citep[see e.g.][]{Kni11}.

One way that long-term variability can be induced is through
irradiation of the donor star from the accreting WD that heats the
envelope of the donor star and causes it to expand slightly.
An increase in the mass-transfer rate leads to stronger irradiation
and therefore expansion of the donor star, whereas a decrease leads 
to weaker irradiation and contraction. If the effects are strong
enough, the mass-transfer becomes unstable on long timescales, 
and the system goes through so-called irradiation-induced mass-transfer cycles
\citep[IIMTC,][]{Pod91,Ham93,Kin96,Bun04}. In this theory, the mass-transfer is
through a series of cycles on Myr timescales, with an off-state
where there is little or no accretion, and an on-state, where
the mass-transfer rate slowly increases towards a peak, and then
decreases until returning to the off-state.
\citet{Bun04} find that the parameter space of WD binaries
that are susceptible to IIMTCs is highly uncertain. CVs with relatively massive 
(e.g. $1 M_{\odot}$) main-sequence donors or somewhat evolved donors with 
convective envelopes are most likely to be affected. 
 Giant donors are unlikely to
be affected significantly because the radius variations caused
by the irradiation are small compared to the radial evolution
of the envelope and the reaction to mass loss.

Another way to achieve long term mass-transfer variability is from
episodic mass loss from the binary which can cause cyclic variations of
the Roche-lobe radius. CVs naturally experience such mass loss
events when they erupt as novae \citep{Sha86,Mac86}. 
If the angular momentum loss
is high compared to the mass loss, the orbit contracts in
response to the nova eruption, whereas the orbit widens
if the angular momentum loss is low compared to the mass loss.
The effects of this process are therefore most likely stronger
in systems with extreme mass ratios, where the specific angular
momentum of the two stars is very different.

\subsection{Observations}
\label{sec_ch3:mt_var_obs}

The mass-transfer cycles discussed above are difficult to study
observationally, as the timescales are longer than the time we
have been able to monitor CVs. A useful method is by comparing
systems with similar properties, as they would be expected to
also have similar mass-transfer rates. \citet{Tow09} used the 
effective temperatures of the WDs to trace the mass accretion
rates. In their sample there are seven non-magnetic CVs above the period gap which show a large scatter in WD effective temperatures and inferred mass-transfer rates. This might be evidence for mass-transfer variability.
Below the period
gap, the mass-transfer differences found by \citet{Tow09} are much 
smaller, and a similar result is found by \citet{Pat09} using
time-averaged accretion disk luminosities. The co-existence of
dwarf novae and novae-likes at the same orbital periods adds to
the case of weak mass-transfer variability below the period
gap, but the evidence is not compelling.

The recurrent nova T Pyx might provide evidence for mass-transfer
variability on its own. At a period of 1.83 h \citep{Pat98,Uth10}
it is clearly below the 
period-gap and should therefore be faint with a low mass accretion
rate. However, it is observed as a recurrent nova with a very high
quiescent temperature implying an accretion rate higher than $10^{-8} M_{\odot}$
yr$^{-1}$, two orders of magnitude above ordinary CVs at this period.
Most likely the system is in a transient evolutionary state. 
\citet{Sch10} suggest that it was an ordinary CV until it erupted
as a nova in 1866. This eruption triggered a wind-driven supersoft
X-ray phase, resulting in an unusually high luminosity and accretion
rate \citep{Kni00}. The recurrence time of the nova eruptions of T Pyx has increased,
and \citet{Sch10} argue that the state is not self-sustaining. According to them the mass-transfer rate has decreased from about 10$^{-7} M_{\odot}$ yr$^{-1}$
after the first nova eruption in 1866 to the current rate of about
10$^{-8} M_{\odot}$ yr$^{-1}$. It is therefore likely that it will
cease being a recurrent nova in the near future and return to the
population of faint CVs.

The mass distribution of the white dwarf components in CVs may indicate mass transfer variability as well.  
Contradicting the accepted model of nova eruptions in CVs \citep[see also][]{Zor11b}, white dwarfs in CVs are significantly more massive than single white dwarfs \citep[e.g.][]{War95, Sav11}. If long-term mass-transfer cycles occur in CVs, the masses of the white dwarf components could be significantly enhanced.

\section{Model}
\label{sec_ch3:model}
\subsection{Mass-transfer variability}
From Sect.\,\ref{sec_ch3:mt_var} we conclude that there are both theoretical and observational
support for long-term mass-transfer variability in accreting
WD binaries. 
To understand the effects of the mass-transfer variability on the growth
of white dwarfs we need to model it. However, the observational evidence
hardly constrain the theoretical models which rely on highly uncertain
parameters \citep[e.g.][]{Bun04}. Here our main goal is to understand 
whether the effects of the variability are important, should such 
variability exist, rather than
studying in detail the effects of a particular theoretical model.
We therefore set up a number of models according to the following 
considerations:
The mass-transfer rate cycles between two separate states, on- and off-state,
with a duty cycle $\beta<1$ representing the fraction of the time the source
spends in the on-state. 
In other words 
\begin{equation}
\bar{R}_{\rm MT} = \beta \bar{R}_{\rm on} + (1-\beta)\bar{R}_{\rm off}, 
\label{eq:r_mt}
\end{equation}
where  
$\bar{R}_{\rm MT}$ is the average long-term mass-transfer rate, $\bar{R}_{\rm on}$ the average mass
transfer rate in the on-state, and $\bar{R}_{\rm off}$ the average mass
transfer rate in the off-state. 
In most variability models, the stars do not
fill their Roche-lobe in the off-state, either because the stars shrink,
or the orbit expands, and binary models show dramatic drops in the mass-transfer
rate when this happens. We therefore assume that all of the accretion takes place in
the on-state. Even if mass-transfer in the off-state only drops by a factor of $f\approx 10$  below the average mass-transfer rate, i.e. 
\begin{equation}
\bar{R}_{\rm off} = \bar{R}_{\rm MT} /f,
\label{eq:r_off}
\end{equation}
the fraction of mass transferred in this state is $(1-\beta)/f$ and therefore only changes our results at the percentage level.
The behaviour in the on-state is probably different from system to system,
but to retain the {\it average} long-term mass-transfer rate $\bar{R}_{\rm MT}$, the average mass
transfer rate in the on-state must be 
\begin{equation}
\bar{R}_{\rm on}=\frac{\bar{R}_{\rm MT}}{\beta}\cdot(1-\frac{1-\beta}{f})\simeq\bar{R}_{\rm MT}/\beta. 
\label{eq:r_on}
\end{equation}

\begin{figure*}
\centering
\begin{tabular}{cc}
\includegraphics[angle=0, width=\columnwidth]{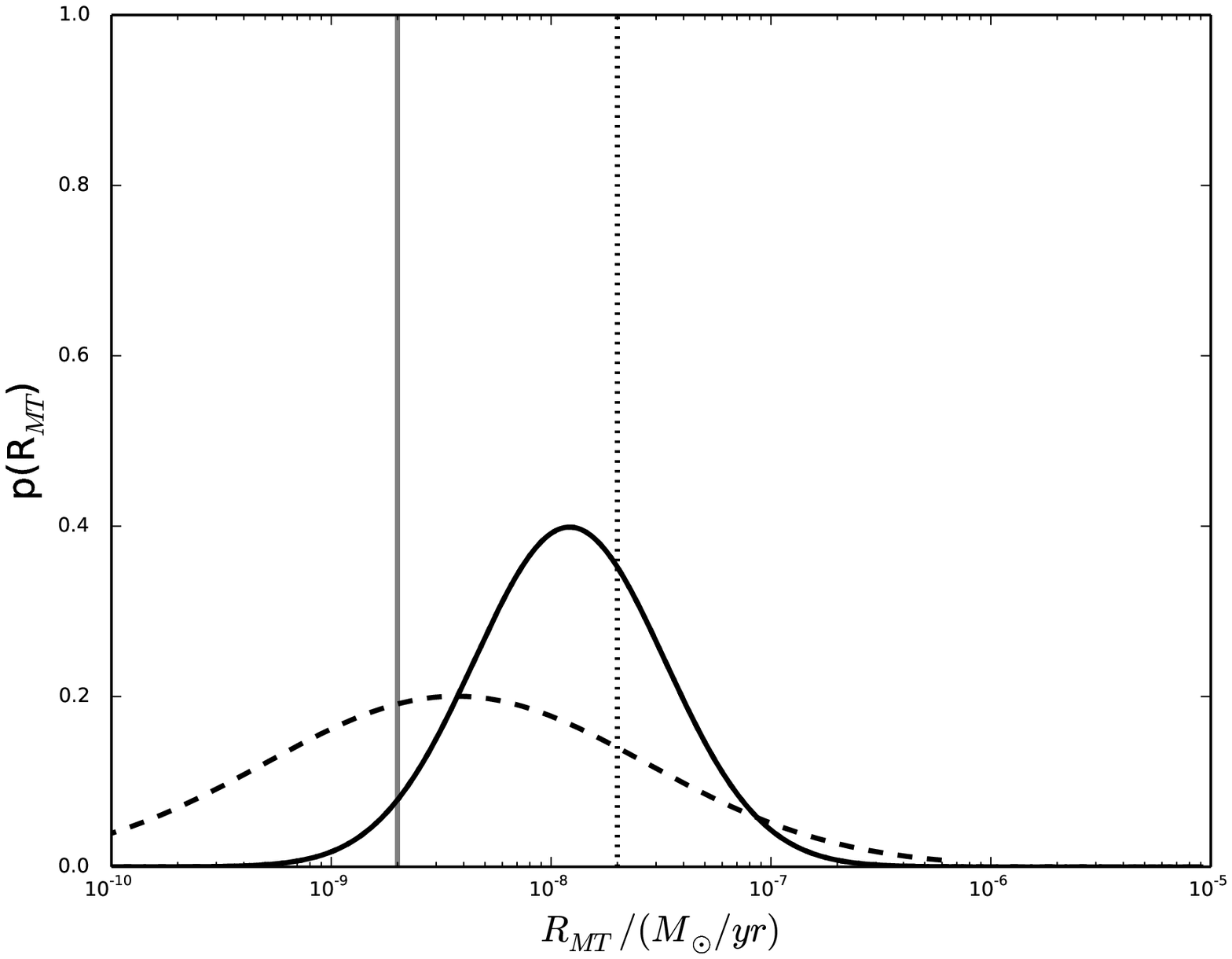}
\includegraphics[angle=0, width=\columnwidth]{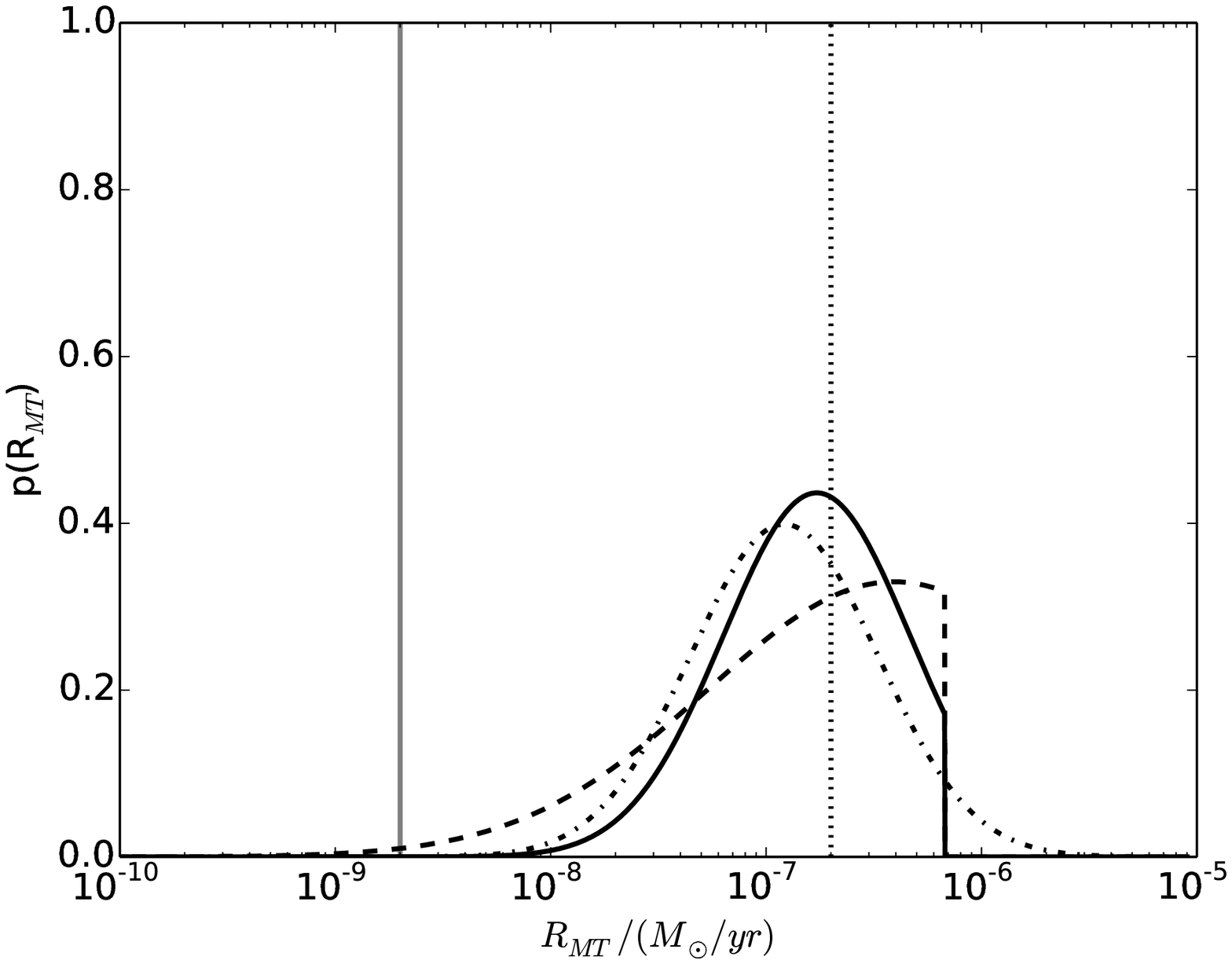}
\end{tabular}
\caption{Example of the mass-transfer variability models, for an
average mass-transfer rate of $\bar{R}_{\rm MT}=2.0\times10^{-9}M_{\odot}$ yr$^{-1}$
(grey line). The lines show the fraction of time that the system spends with a mass-transfer rate between $R_{\rm MT}$ and $R_{\rm MT} + \mathrm{d} R_{\rm MT}$ in the on-state for each of the 
models. The black lines indicate models with duty cycles of $\beta=0.1$ on the left and $\beta=0.01$ on the right. 
The dotted line is model CONST, the dash-dotted line is model 
NORM with $\sigma_e=1$, and the solid line model NORM-MAX
with $\sigma_e=1$. The dashed line is also model NORM-MAX with
with a larger spread $\sigma_e=2$.
}
\label{fig_ch3:MTmodels}
\end{figure*}

We employ two
model types: (model CONST) a constant mass-transfer rate, representing systems that 
quickly
attain and keep their peak rate, and (model NORM) a lognormal probability distribution (with a standard deviation $\sigma_e$ given in base {\it e}),
representing systems with a gradual increase (or decrease) of the mass
transfer rate, as is typically seen in the IIMTC scenario. Examples of
these models are shown in Fig.\,\ref{fig_ch3:MTmodels}, 
showing the fraction of time that
a system with an average mass-transfer rate of $\bar{R}_{\rm MT}=2.0\times10^{-9}M_{\odot}$ yr$^{-1}$  
spends at different
accretion rates.

Also shown in Fig.\,\ref{fig_ch3:MTmodels} is a model (model NORM-MAX), 
in which there is a maximum accretion rate that the systems can reach. The reason
for this is that the models for hydrogen surface burning have a critical mass
$\dot{M}_{\rm crit}$, above which mass is accreted too fast. The luminosity of 
the hydrogen burning at $\dot{M}_{\rm crit}$
is similar to the Eddington luminosity, and it is normally assumed that the surplus is ejected. 
At first the surplus piles up on the white dwarf as a ``red giant'' envelope, however, the envelope may interact with the binary through a common-envelope (CE) phase \citep{Pac76} or a wind may develop from the envelope \citep{Hac96}.
The 
density of the envelope material is high enough to obscure the X-rays from
the hydrogen burning. As this irradiation is necessary to keep the
mass-transfer rate high in the IIMTC scenario, the rate is unlikely
to exceed $\dot{M}_{\rm crit}$ for long. For our lognormal model we therefore
redistribute the parts of the probability density function above
$\dot{M}_{\rm crit}$ to lower mass-transfer rates, modifying the function to
retain the average mass-transfer rate. For model NORM-MAX it is per definition not possible to construct
models with $\beta\leq\bar{R}_{\rm MT}/\dot{M}_{\rm crit}$ as
too much time is spent at low accretion rates to reach $\bar{R}_{\rm MT}$.
For these models we therefore gradually increase the duty cycle
so $\beta=\bar{R}_{\rm MT}/\dot{M}_{\rm crit}$ when necessary.

We furthermore assume that for mean mass-transfer rates in the classical
hydrogen burning regime (higher than a few times$10^{-7}M_{\odot}$yr$^{-1}$) the
variability disappears. In almost all systems with such high mass
transfer rates, the mass-transfer is transferred on the thermal
timescale of the donor star, which is shorter than or comparable to
the timescale of the mass-transfer cycles (i.e. the star does not
have the time to adjust to the heating before the heated layers are
lost). 

\subsection{Integrated retention efficiency}
\label{sec_ch3:ret}
\begin{figure*}
\centering
\begin{tabular}{cc}
\includegraphics[angle=0, width=\columnwidth]{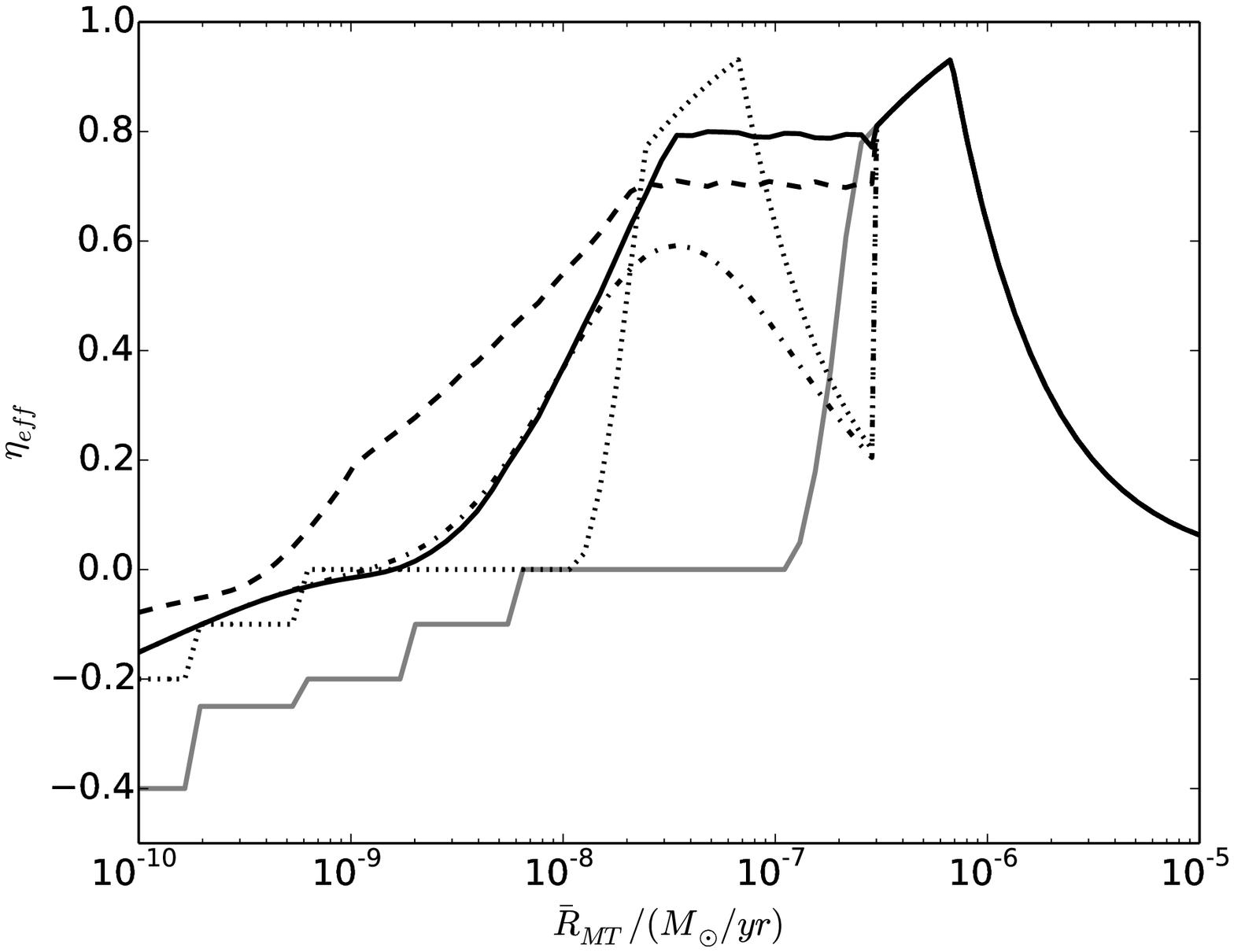} &
\includegraphics[angle=0, width=\columnwidth]{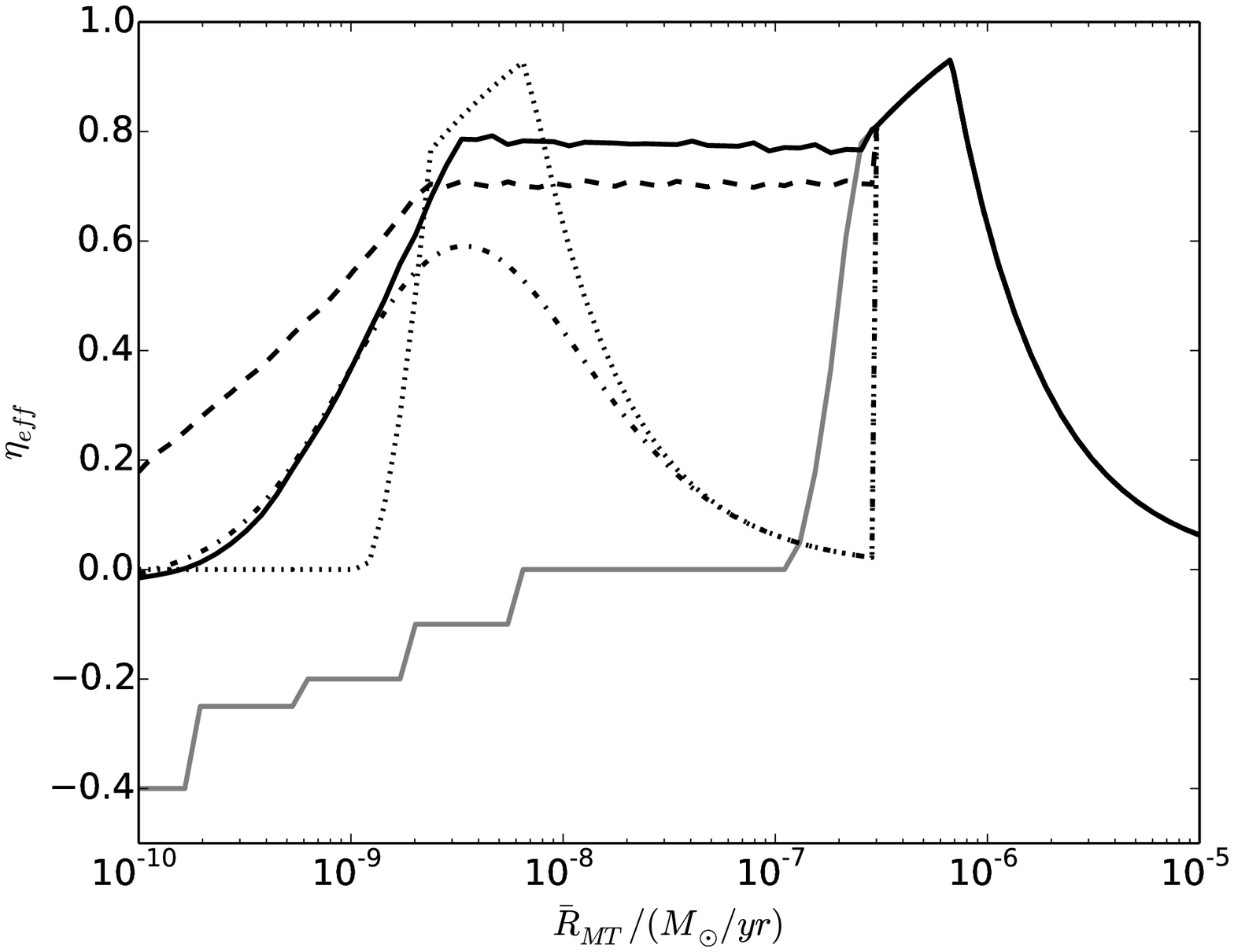} \\
\end{tabular}
\caption{Effective retention efficiency as a function of average mass-transfer rate for
different mass-transfer models for a 1.3\Msolar~WD accretor. The grey line shows a model without mass-transfer variability.
Model CONST is shown as a dotted line and model NORM with $\sigma_{e}=1$ is shown as the dash-dotted line. The solid black and dashed lines are model NORM-MAX with $\sigma_{e}=1$ and $\sigma_{e}=2$ respectively. On the left models with a duty cycle of $\beta=0.1$ are shown and on the right $\beta=0.01$. }
\label{fig_ch3:effective}
\end{figure*}

The retention efficiency $\eta$ is the fraction of mass transferred that is
retained by the WD. This is the fraction of hydrogen that is burned stably
into helium \textit{and} where the helium is also burned stably.
The fraction of mass $\eta$ that is retained depends on the mass of 
the WD and on the accretion
rate. 
We estimate $\eta$ based on \citet{Hac08} for hydrogen burning and \citet{Kat99} for helium burning. It is the same prescription as the optimistic case in \citep{Bou13}, see their Eq.5 and A1-A5. We assume that the wind-stripping effect \citep{Hac99b} is not effective, i.e. $c_1=0$.
We make one adjustment at low accretion rates where we assume that the retention factor is $\eta\leq$0, corresponding
to a net loss of mass from the white dwarf, with values estimated from
\citet{Pri95}.
The model is shown in Fig.\,\ref{fig_ch3:effective},
where the final retention efficiency as a function of the mass-transfer
rate is shown as the grey line. We use this
model in the following analysis, but we caution that
the theoretical models that this is based on are calculated assuming
constant accretion rates. As discussed above, the properties of the 
WD (in particular
the temperature) depend on the accretion history. Therefore it is not clear
if these models are accurate for systems with variable accretion
rates. However, we note that in the irradiation-induced mass-transfer
scenario, the change in the mass-transfer rate is slow enough
(timescales of Myr) that the assumption of a constant mass-transfer
rate is likely to be justified.

If we know the retention efficiency as a function of the
mass-transfer rate, $R_{\rm MT}$, we can find the effective retention factor for a 
given {\it mean} mass-transfer rate $\bar{R}_{\rm MT}$ for each of the models:
\begin{equation}
\label{eq_ch3:reten}
\eta_{eff}=\frac{\int_0^{\infty}p(R_{\rm MT})\cdot R_{\rm MT}\cdot \eta (R_{\rm MT})dR_{\rm MT}}{\bar{R}_{\rm MT}}
\end{equation} 
where $p(R_{\rm MT})$ is the model probability of a given mass-transfer rate
$R_{\rm MT}$.

Fig.\,\ref{fig_ch3:effective} shows the results of applying Eq.\,\ref{eq_ch3:reten}
to the examples of the mass-transfer models. By construction, all of the models 
conform to the shape given by the grey line in Fig. \ref{fig_ch3:effective} in the stable burning regime (with mass-transfer rates of a few times
$10^{-7}M_{\odot}$ to $\dot{M}_{\rm crit}$), where we assume that there is no 
variability. Below this range the models with mass-transfer
cycles clearly distinguish themselves from the model without, as they
are able to retain a significant fraction of the accreted mass at
much lower mean accretion rates of about $\beta \cdot 10^{-7}M_{\odot}$yr$^{-1}$,
irrespective of the details of the model. The differences between the
models are easily understood: model CONST corresponds to a simple
shift of the average mass-transfer rate by a factor 1/$\beta$, and
the retention curve is therefore keeping its narrow shape, whereas
model NORM both shifts the curve and broadens it due to the
lognormal variability. The maximum is shifted slightly downwards for
this model, due to the difference between the mean and the median of
a lognormal model. Both curves display local minima in the retention
curves near $\bar{R}_{\rm MT}\simeq10^{-7}M_{\odot}$ yr$^{-1}$, because above
this value the peaks of the mass-transfer cycles are located above
$\dot{M}_{\rm crit}$. As we argue above, this is probably not realistic
due to the obscuration of the X-rays from the white dwarf surface
at these high accretion rates. Most likely the systems that have
accretion rates in this range (approximately $10^{-8}$ to $10^{-7}$ in Fig.\,\ref{fig_ch3:effective}) have higher duty cycles (the
depressions only appear for duty cycles $\beta\lesssim0.1$)
and/or lower peak accretion rates than assumed, and therefore also retain 
much of the accreted mass.
Indeed in model NORM-MAX where
accretion is not allowed to exceed $\dot{M}_{\rm crit}$, 
the retention efficiency stays high in this
accretion range.

For models NORM and NORM-MAX the retention efficiency
stays above zero well below $\bar{R}_{\rm MT}=10^{-9}M_{\odot}$ yr$^{-1}$,
despite the systems spending more time in accretion states with
negative retention efficiencies, because even if they only spend a
short time at high $R_{\rm MT}$, the fraction of mass accreted in this
regime is still considerable. We believe that model NORM-MAX captures the behaviour of IIMTCs best, such as the ones
modelled by \citet{Bun04}, because the formation of a WD envelope at high mass transfer rates is likely to quench the irradiation process (see also  Sect.\,\ref{sec_ch3:model}). We conclude that for all of the
models the WDs can effectively grow down to average mass-transfer
rates a factor of $\beta$ lower than in the standard scenario without
variability, irrespective of the specific shape of the variability
(assuming that the mass-transfer rate does not exceed $\dot{M}_{\rm crit}$).

\begin{figure*}
\centering
\includegraphics[width=\textwidth]{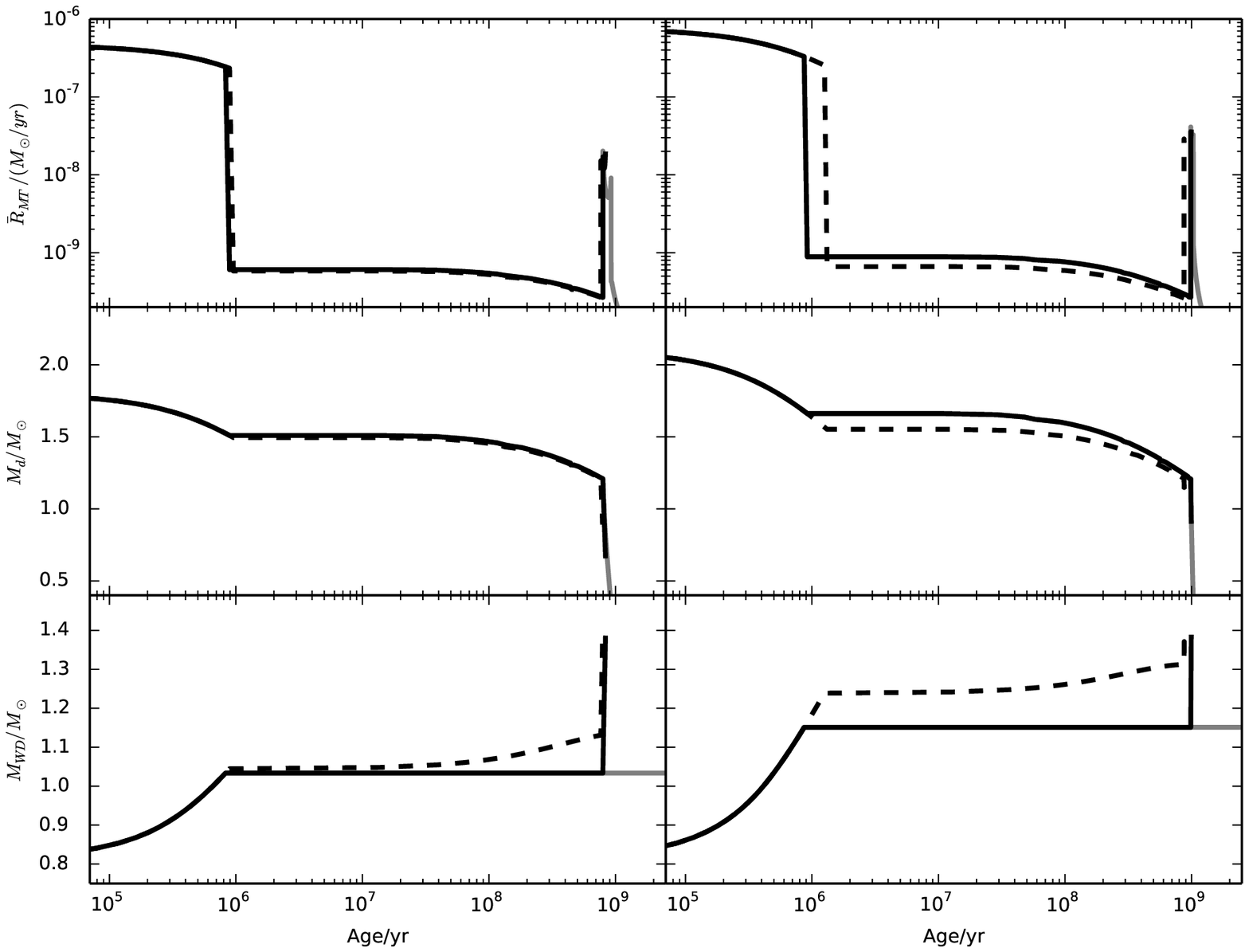}
\caption{Binary tracks for two WD binaries starting mass transfer while the
donor star is on the main sequence. The initial mass of the WD is 
0.8\Msolar, the initial orbital separation is 4.5\Rsolar, and the initial mass of the donor star is 1.8\Msolar~(left) and 2.1\Msolar~(right). The panels show the evolution
of the (time-averaged) mass-transfer rate $\bar{R}_{\rm MT}$ (upper), the donor mass $M_d$ (middle) and the
WD mass $M_{WD}$(lower), for three different models: SeBa standard (grey solid),
model NORM-MAX with $\beta=0.1$~and~$\sigma_e=1$ (black solid) and model NORM-MAX with
$\beta=0.01$~and~$\sigma_e=1$ (black dashed). }
\label{fig_ch3:startracks}
\end{figure*}

\section{Application to binary stellar evolution}
\label{sec_ch3:binevol}
Our models of mass-transfer cycles from Sect.\,\ref{sec_ch3:model} significantly modify and enhance the mass retention efficiency of accreting white dwarfs (see Fig.\,\ref{fig_ch3:effective}). This can have a significant effect on the characteristics of the population of accreting white dwarf binaries, e.g. the distribution of WD masses in cataclysmic variables. Furthermore, the growth of carbon-oxygen white dwarfs is important for understanding the rate of SNIa and their delay-time distribution in the single-degenerate channel, which we study here as an example of the implications of mass-transfer variability. 

In the traditional picture without variability, the systems
that can become type Ia supernovae are distributed in
two regions (``islands'') in the plane of the 
two parameters - orbital period and secondary mass - just after 
the formation of the WD \citep{Li97, Hac99b, Han04}. 
One of the islands consists of progenitors 
where the companion has evolved to a giant before commencing the mass
transfer. As mentioned above, these systems are not likely to
be susceptible to IIMTCs \citep{Bun04}.
The other island consists of main-sequence or slightly evolved
donors. If the mass of the donor star is much higher than the
white dwarf mass, the mass transfer is dynamically or tidally unstable, leading to a merger of the two stars,
leading to a natural upper mass limit to the island. The lower
mass limit is determined by the fact that when the mass of the
donor star comes near to the mass of the white dwarf, the mass
transfer rate drops below the stable surface hydrogen burning
limit. This typically happens around 1.5\Msolar, and since
the vast majority of CO WDs are born below 1\Msolar, the initial
mass of the donor star must be above approximately $2 M_{\odot}$.

In the models with mass-transfer variability it is possible to
retain the matter accreted at lower mass-transfer rates. This
does not affect the upper limit to the donor mass, since this
limit is determined by the stability of mass transfer at high rates.
However, the lower mass limit of the donor star is likely to be 
affected. 
Therefore an increased retention at low mass-transfer rates allows
WDs in binaries with lower donor masses to grow, and therefore allows
systems with lower initial donor masses to become type Ia
supernovae. 

The limits depend on the strengths and shape of the mass-transfer
variability, but also on the evolution of the donor star and its
reaction to the mass loss.
To better understand what our results mean for the evolution of
binaries with WDs, we have calculated binary evolutionary sequences
with the binary population synthesis code SeBa \citep{Por96, Nel01, Too12, Too13}. Our goal is to understand how the possible 
long-term variability affects the evolution of accreting WD
binaries that might become type Ia supernovae. We therefore
compare evolutionary tracks computed with a standard SeBa model
to tracks 
where the accretion efficiency has been modified by variability. 
For the model without mass transfer variability, the retention efficiency $\eta$ as depicted by the grey line in Fig.\,\ref{fig_ch3:effective} is adopted in SeBa (see also Sect.\,\ref{sec_ch3:ret}). For model NORM-MAX, the standard retention efficiency is additionally modified to
\begin{equation}
\eta = \left\{ 
\begin{array}{rl}
0.8 &\mathrm{~ if ~ ~} \beta \dot{M}_{\rm ST} < \bar{R}_{\rm MT} < \dot{M}_{\rm ST} \\
0.8(\mathrm{log}(\bar{R}_{\rm MT}) -\mathrm{log}(\beta\dot{M}_{\rm ST})) &\mathrm{~ if ~ ~} 0.1\beta \dot{M}_{\rm ST} < \bar{R}_{\rm MT} < \beta\dot{M}_{\rm ST} \\
0 &\mathrm{~ if ~ ~}  \bar{R}_{\rm MT} < 0.1\beta\dot{M}_{\rm ST} \\
\label{eq_ch3:ret_mtvar}
\end{array} \right. 
\end{equation}
where 
\begin{equation}
\bar{R}_{\rm MT} < \dot{M}_\mathrm{st} = 3.1 \cdot 10^{-7} \Big(\frac{M_\mathrm{WD}}{\mathrm{M}_{\odot}}-0.54\Big). 
\label{eq_ch3:Mst}
\end{equation}

As can be seen from
Fig.\,\ref{fig_ch3:effective}, other variability models might give
somewhat smaller effects if the peak accretion rate is not limited. 
For each model, we make the simplifying assumption that the effective 
retention $\eta_{eff}$ only depends on $\bar{R}_{MT}$ and the mass 
of the WD, i.e. that the shape and strength of the IIMTCs are the same 
irrespective of the properties of the donor star. This is clearly
unrealistic. However, the goal of our study is to understand \textit{if}
the variability is likely to impact the SNIa population properties, 
and to indicate what the possible effects might be, and the
assumption is sufficient for this. 

In Fig.\,\ref{fig_ch3:startracks} we show the results of evolving
two WD binaries with close main-sequence companions according 
to the standard SeBa model (grey solid lines) and model NORM-MAX 
with $\sigma_e=1$, and $\beta=0.1$ (black solid lines) and $\beta=0.01$ 
(black dashed lines). The systems behave similarly after the initial contact,
when the mass-transfer rate is high. When it drops below the
standard surface burning regime, differences appear, not just in
the WD growth (bottom panel), but also in the time-averaged mass-transfer rate itself
(top panel). This is because the matter ejected from the system 
carries angular momentum, which can strongly affect the evolution of the binary orbit.  
We assume that the matter that can not be accreted by the WD leaves the system with the specific orbital angular momentum of the WD. Note that this is the only way in which we allow the time-averaged mass transfer rate to vary in our models. 

The main point of Fig.\,\ref{fig_ch3:startracks} is that for both systems of the standard model the mass of the WD 
never reaches the critical explosion mass (approximately $1.4 M_{\odot}$, for
this model the initial companion mass must be about 2.3\Msolar~for the WD to reach this mass), whereas the variability models
do reach the explosion mass.

\section{Binary population Synthesis}
\label{sec_ch3:bps}
In the previous sections we have shown that mass-transfer variability has
the potential to significantly change the parameter space of initial
WD binaries that can become type Ia supernovae, towards both lower-mass
donor stars as well as lower-mass white dwarfs. To understand how this
can affect the population of type Ia supernovae, we 
use the binary population synthesis (BPS) code SeBa to model the evolution of SNIa progenitors according to different mass-transfer variability models.

In SeBa, stars are evolved from the zero-age main sequence (ZAMS) until remnant formation, and, at every timestep, processes such as stellar winds, mass transfer, angular momentum loss, magnetic braking and gravitational radiation are taken into account with appropriate recipes. 
Magnetic braking \citep{Ver81} is based on \citet{Rap83}.
The initial stellar population is generated with a Monte-Carlo approach according to appropriate distribution functions. 
Initial primary masses are drawn from 0.95-10\Msolar~from a Kroupa IMF \citep{Kro93} and secondary masses from a flat mass ratio distribution between 0 and 1. The semi-major axis of the binary is drawn from a power
law distribution with an exponent of -1 \citep{Abt83}, ranging from 0 to 10$^6$\Rsolar~and the eccentricity
from a thermal distribution, ranging from 0 to 1 \citep{Heg75}.
Furthermore, solar metallicities are assumed. For the normalization of the simulation, a binary fraction of 50\% is assumed and an extended range of primary masses between 0.1-100\Msolar.

The CE-phase \citep{Pac76} plays an essential role in binary evolution in the formation of close binaries with compact objects. 
Despite of the importance of the CE-phase and the enormous efforts of the community, we still do not understand the phenomenon in detail. To take into account the uncertainty in the CE-phase in our models, we differentiate between two CE-models.
The canonical CE-formalism is the $\alpha_{\rm CE}$-formalism \citep{Tut79, Web84} that is based on the energy budget of the binary system. 
The $\alpha_{\rm CE}$-parameter describes the efficiency with which orbital energy $E_{\rm orb}$ is consumed to unbind the CE, i.e.
\begin{equation}
\frac{G M_{\rm 1} M_{\rm 1, e}}{\lambda R} = \alpha_{\rm CE} (E_{\rm orb,init}-E_{\rm orb,final}),
\label{eq_ch3:alpha-ce}
\end{equation}
where $M_{\rm 1}$, $M_{\rm 1,env}$ and $R$ are the mass, envelope mass and radius of the donor star and $\lambda$ is the envelope structure parameter \citep{DeK87}. Based on the evolution of double WDs, \citet{Nel00} derives a value of $\alpha_{\rm CE}\lambda=2$, which we have assumed here.  

An alternative CE-prescription was introduced by \citet{Nel00} in order to explain the observed distribution of double WDs systems. The $\gamma$-formalism of CE-evolution is based on the angular momentum balance. The $\gamma$-parameter describes the efficiency with which orbital angular momentum is used to expel the CE according to: 
\begin{equation}
\frac{J_{\rm b, init}-J_{\rm b,  final}}{J_{\rm b,init}} = \gamma \frac{\Delta M_{\rm 1}}{M_{\rm 1}+ M_{\rm 2}},
\label{eq_ch3:gamma-ce}
\end{equation} 
where $J_{\rm b}$ is the orbital angular momentum of the binary, and $M_{\rm 2}$ is the mass of the companion. We assume $\gamma=1.75$ \citep{Nel01}. Although assuming the $\gamma$-prescription in BPS codes leads to a significant improvement in the synthetic double WD population, the physical mechanism remains unclear. Recently \citet{Woo10_Myk, Woo12} proposed that double WDs can be formed by stable, non-conservative mass transfer between a red giant and a main-sequence star. The effect on the orbit is a modest widening, with a result not unlike the $\gamma$-description. For a review on CE-evolution, see \citet{Web08} and \citet{Iva13}.

\begin{table}
\centering 
\caption{Time-integrated SNIa rates in the SD channel for different mass-transfer variability models and the common envelope prescriptions in units of $10^{-4}$\Msolar$^{-1}$.}
\begin{tabular}{lcc}
\thickhline
 & $\gamma$-prescription & $\alpha_{\rm CE}$-prescription\\
No variability					& 0.59 	&  0.79\\
Model NORM-MAX ($\beta=0.1$) 		& 1.2 	& 1.6 \\
Model NORM-MAX ($\beta=0.01$)		& 1.4 	& 2.0 \\
\\
Observed & \multicolumn{2}{c}{$4->34^1$}\\
\thickhline
\hline
\end{tabular}
$^1$ \citet{Mao10, Gra13}, see also Sect.\,\ref{sec_ch3:dis} for a discussion on the observed rates. 
\label{tbl_ch3:int_rate}
\end{table}

\section{Results and discussion}
\label{sec_ch3:dis}
\begin{figure*}
    \centering
    \setlength\tabcolsep{0pt}
    \begin{tabular}{ccc}
	\includegraphics[width=0.35\textwidth]{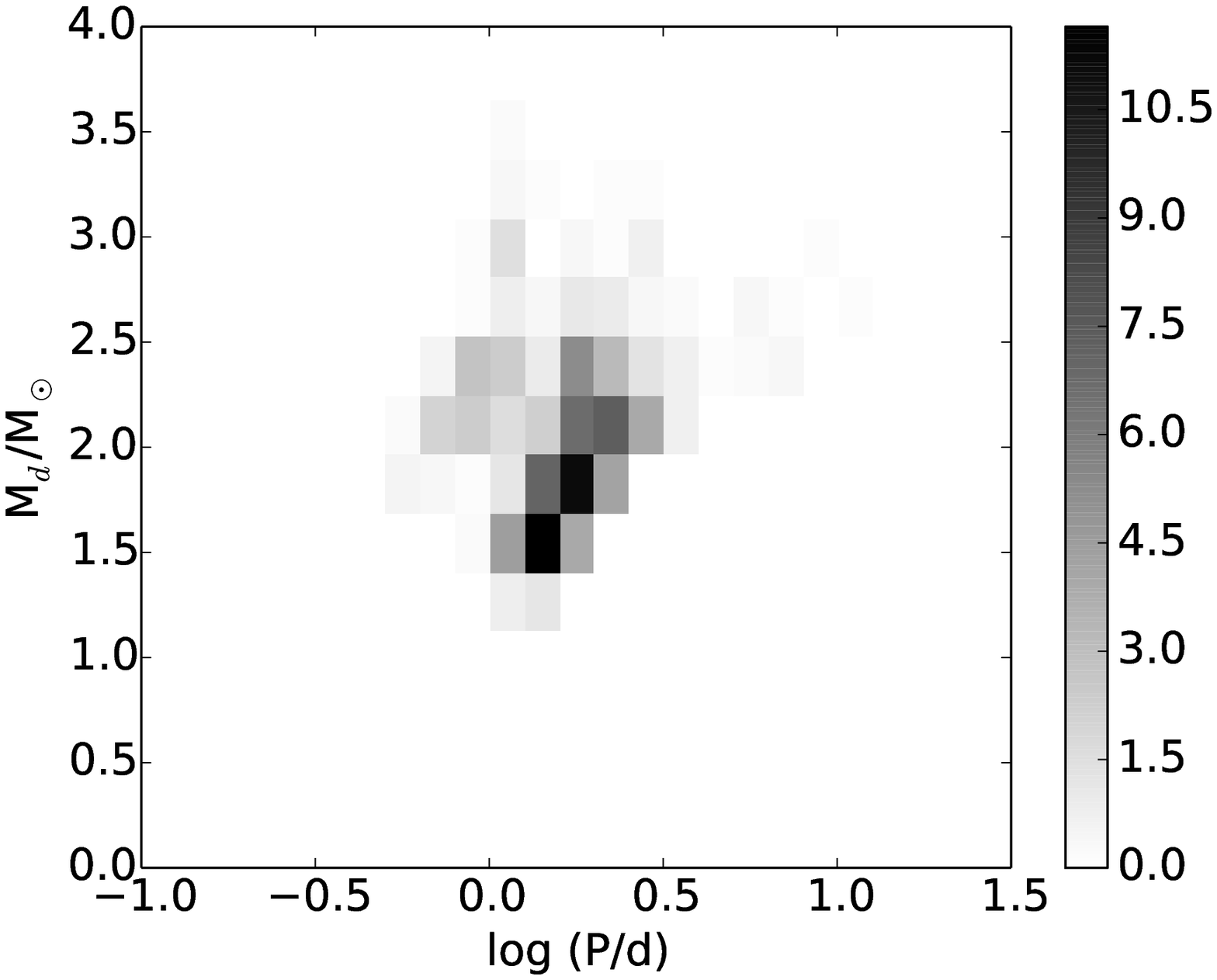} &
	\includegraphics[width=0.35\textwidth]{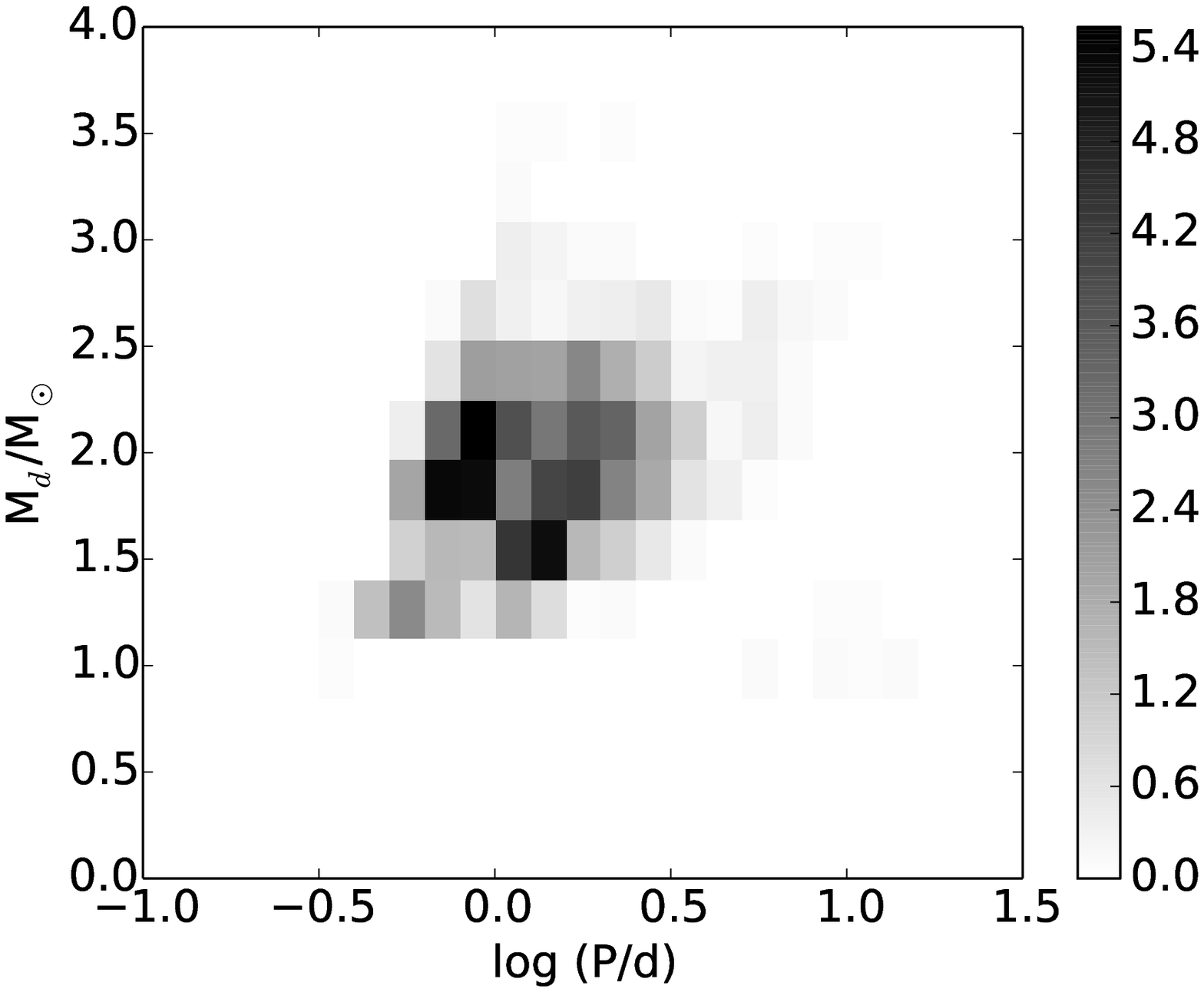} &
	\includegraphics[width=0.35\textwidth]{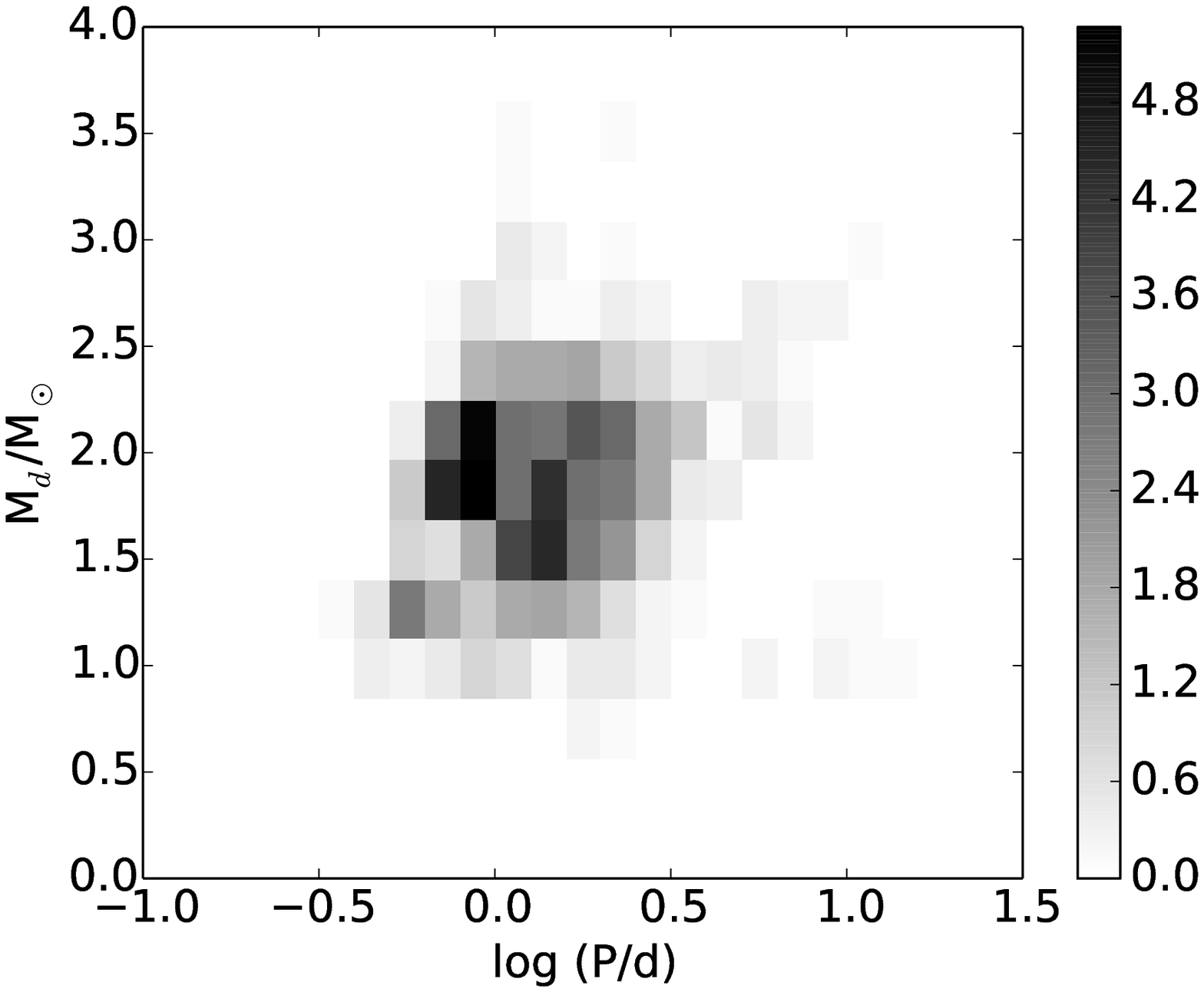} \\
	(a) & (b) & (c) \\
	\end{tabular}
	\caption{Donor mass vs. orbital period after the last mass-transfer event in which the WD is formed for the SD SNIa progenitors assuming the $\gamma$-algorithm with $\gamma=1.75$ for three different mass-transfer models. 
On the left a model without mass-transfer variability, in the middle model NORM-MAX with $\beta=0.1$ and on the right model iii with a duty cycle of $\beta=0.01$. The intensity of the grey scale corresponds to the density of objects on a linear scale in percentages of all SD SNIa progenitors of the corresponding model.
}
    \label{fig_ch3:isl_gamma}
    \end{figure*}

\begin{figure*} 
    \centering
    \setlength\tabcolsep{0pt}
    \begin{tabular}{ccc}
	\includegraphics[width=0.35\textwidth]{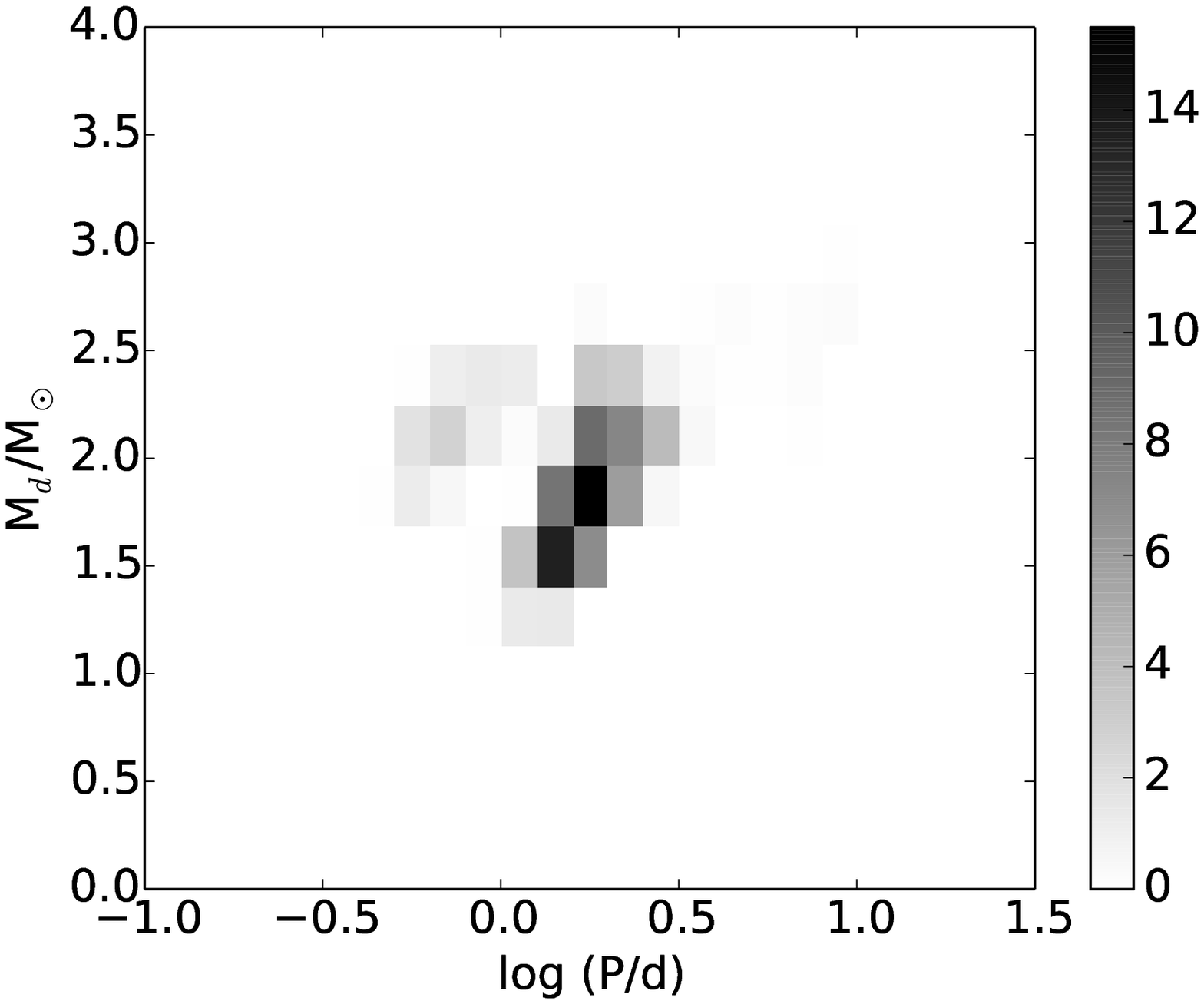} &
	\includegraphics[width=0.35\textwidth]{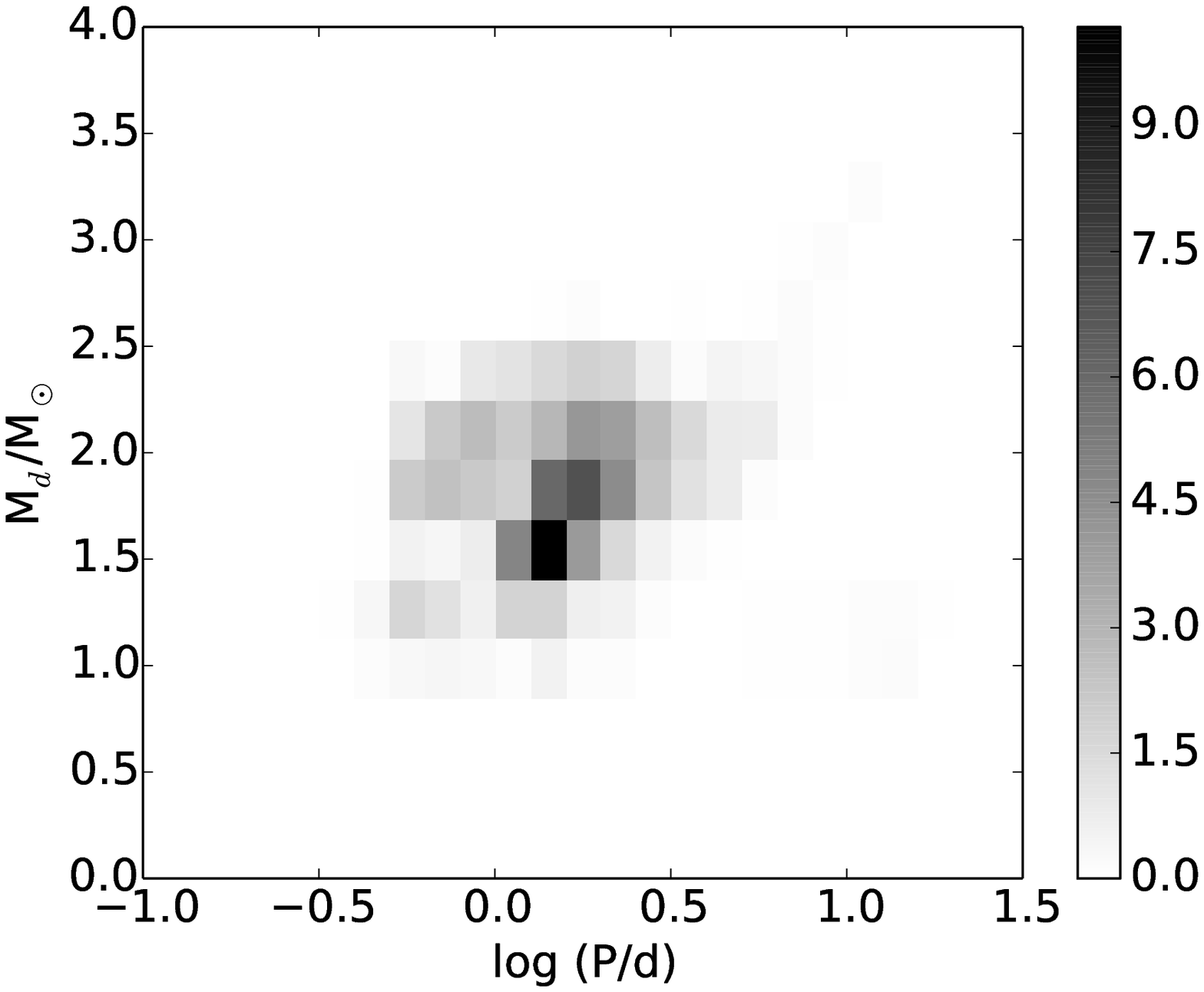} &
	\includegraphics[width=0.35\textwidth]{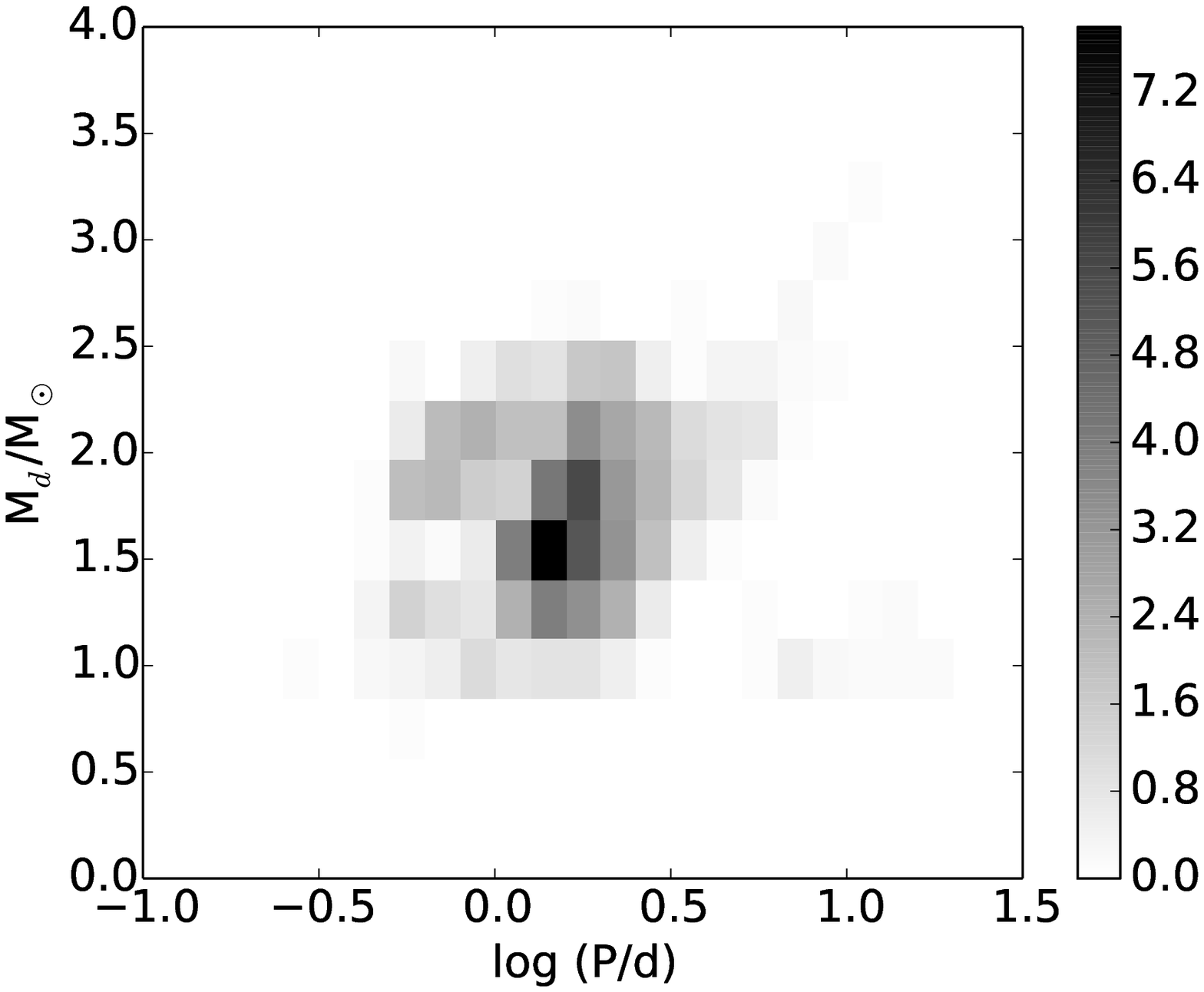} \\
	(a) & (b) & (c) \\
	\end{tabular}
	\caption{Donor mass vs. orbital period after the last mass-transfer event in which the WD is formed for the SD SNIa progenitors assuming the $\alpha_{\rm CE}$-algorithm with $\alpha_{\rm CE}\lambda=2$. 
On the left a model without mass-transfer variability, in the middle model NORM-MAX with $\beta=0.1$ and on the right model NORM-MAX with a duty cycle of $\beta=0.01$. The intensity of the grey scale corresponds to the density of objects on a linear scale in percentages of all SD SNIa progenitors of the corresponding model.}
    \label{fig_ch3:isl_alpha}
    \end{figure*}

Fig.\,\ref{fig_ch3:isl_gamma}~and~\ref{fig_ch3:isl_alpha} shows the systems that become type Ia supernovae in the diagram of 
orbital period - donor mass at the birth of the WD according to SeBa. Fig.\,\ref{fig_ch3:isl_gamma}a~and~\ref{fig_ch3:isl_alpha}a shows the distribution of classical SD SNIa progenitors. Most systems have low-mass donor stars and relatively long periods in accordance with \citet{Hac08} and \citet{Cla14}. Fig.\,\ref{fig_ch3:isl_gamma}b~and~\ref{fig_ch3:isl_alpha}b show how the parameter space of systems that can become type Ia supernovae is extended for model NORM-MAX with a duty cycle of $\beta=0.1$, and Fig.\,\ref{fig_ch3:isl_gamma}c~and~\ref{fig_ch3:isl_alpha}c for a lower duty cycle of $\beta=0.01$. These four figures show that the  parameter space of SNIa progenitors extends to lower donor masses when mass-transfer cycles are taken into account.

\begin{figure*}
    \centering
    \begin{tabular}{cc}
	\includegraphics[width=0.5\textwidth]{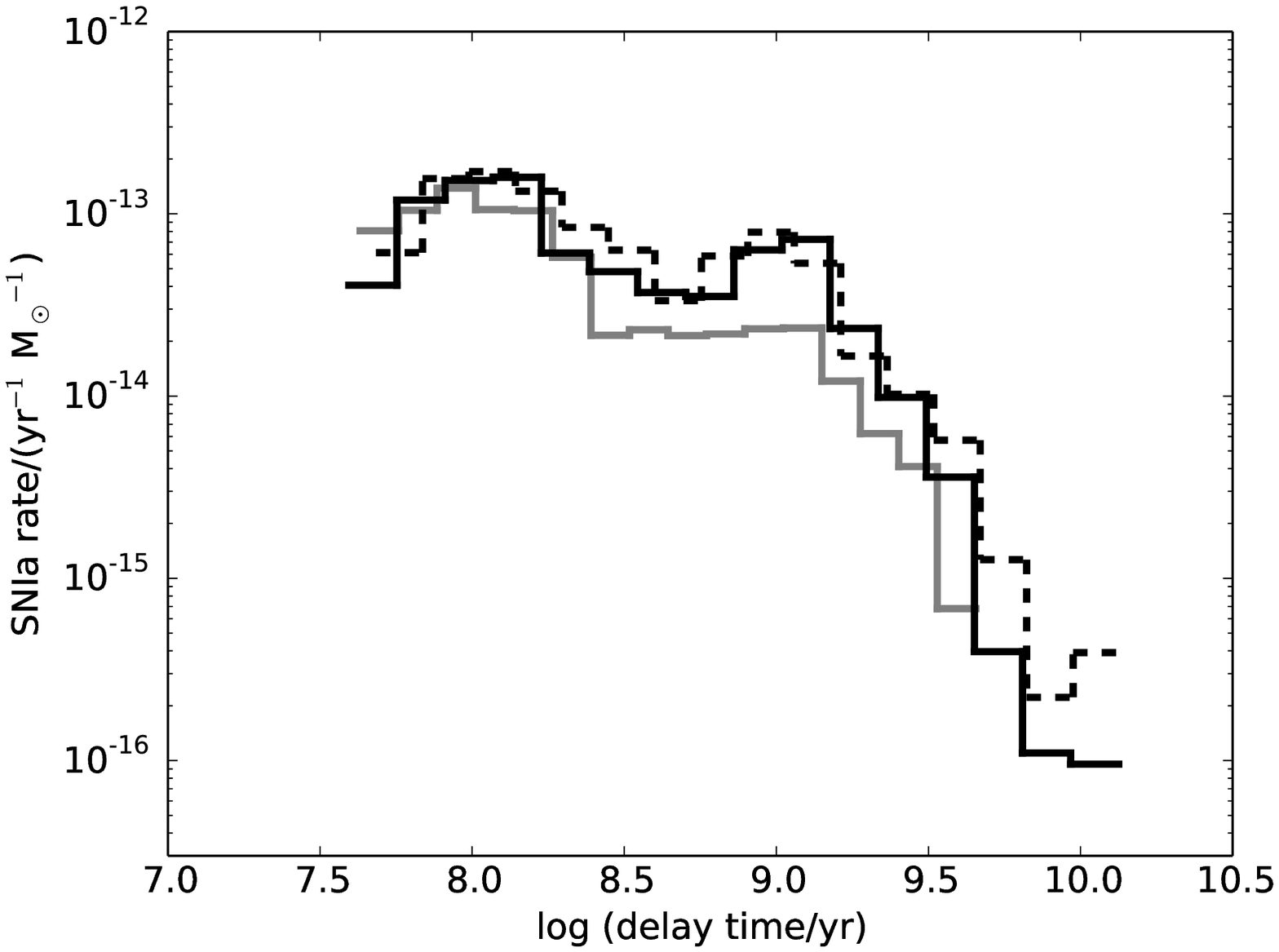} &
	\includegraphics[width=0.5\textwidth]{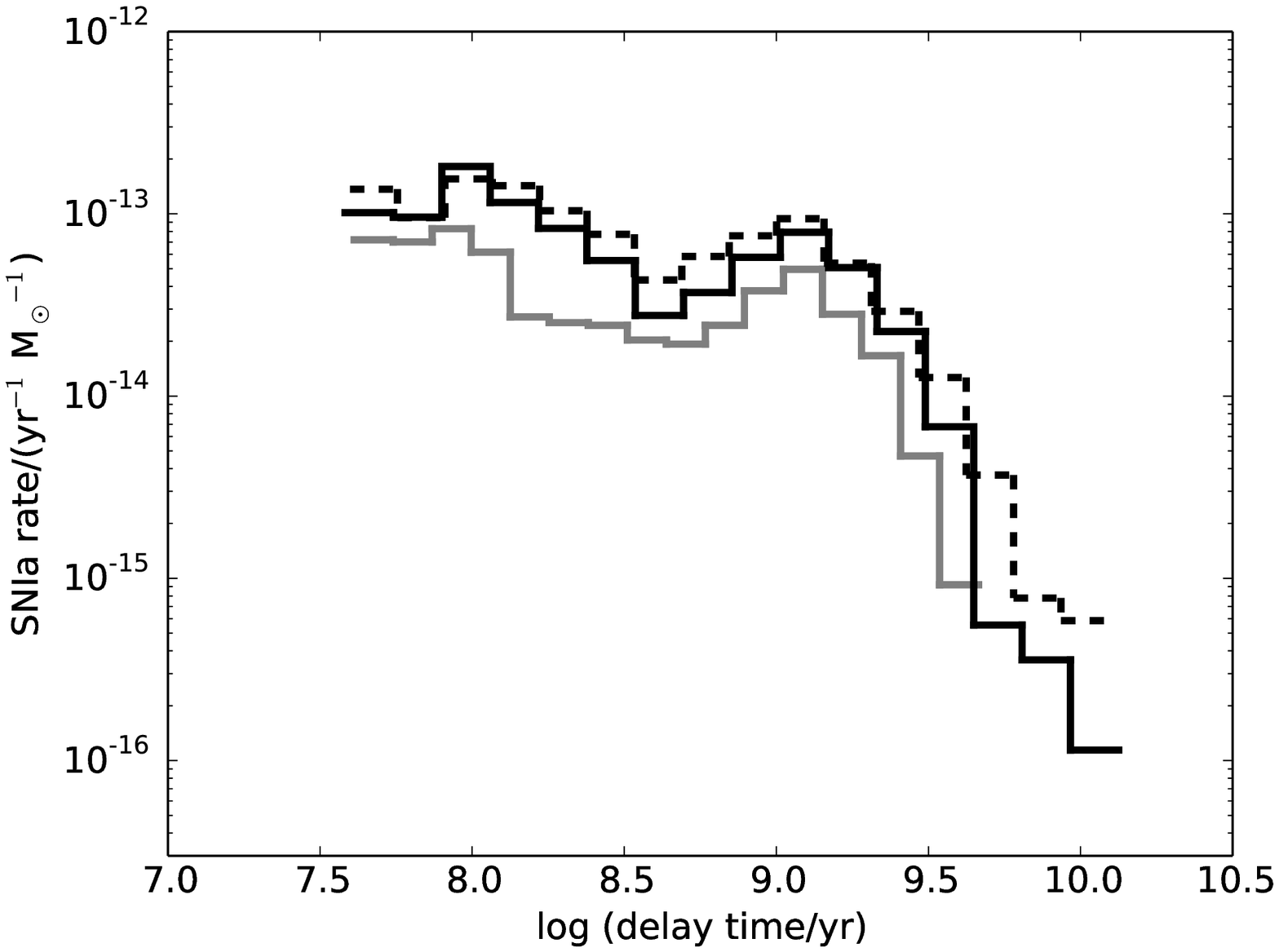} \\
	\end{tabular}
	\caption{Delay time distribution of SNIa events from the SD channel for three different mass-transfer models. The black lines indicate model NORM-MAX with $\beta=0.1$ (solid) and $\beta=0.01$ (dashed). The grey line shows a model without mass-transfer variability. 
On the left assuming the $\gamma$-algorithm with $\gamma=1.75$ and on the right assuming the $\alpha_{\rm CE}$-algorithm with $\alpha_{\rm CE}\lambda=2$.}
    \label{fig_ch3:dtd}
    \end{figure*}

The time-integrated number of SNIa events is about $10^{-4}$\Msolar$^{-1}$, see Table\,\ref{tbl_ch3:int_rate}. When taking into account mass-transfer variability according to model NORM-MAX, the rate is increased by a factor of 2 and 2.5 for $\beta=0.1$ and $\beta=0.01$ respectively, compared to the standard model of non-variable mass-transfer rates. The integrated rates of Table\,\ref{tbl_ch3:int_rate} are based on approximately 1000-3000 SNIa progenitors in the BPS simulation. The DTDs (assuming a single burst of star formation at $t=0$) from all models show a strong decline with time (see Fig.\,\ref{fig_ch3:dtd}). When mass-transfer variability is included in our simulations, the DTDs are affected at delay times from about 100Myr to a Hubble time. However the shape of the DTDs has not changed significantly. 

The effect of including mass-transfer variability on the SNIa rate is mild, even though the retention efficiency of WD accretion is greatly enhanced in our models. 
The extra systems 
that become SNeIa due to mass-transfer variability is limited, compared to the number of extra ZAMS systems
that are born with secondaries in the extended mass range. Our study
shows that the reason for this is that, as the mass of the secondary
decreases, it becomes harder to create close binaries with massive
WDs. As the initial binary mass ratio is higher for these systems,
the orbital separation is decreased more during the first mass-transfer 
episode from the WD progenitor to the secondary star, and most of the
lower-mass secondaries end up being too close to survive until the
formation of the WD. Most of the systems that do survive experience
Roche-lobe overflow from the WD progenitor (primary) when it has
become a helium star, which significantly increases their donor mass
and therefore decreases the evolutionary timescale. This speed-up
means that they cannot explode as the very delayed supernovae that
could be expected of low-mass secondaries, but rather on relatively
short timescales below 1 Gyr.

From galaxy cluster measurements and cluster iron abundances, \citet{Mao11b} and \citet{Mao10} find an observed integrated rate of $(18-29)\cdot 10^{-4}\ M_{\odot}^{-1}$ and a lower limit of $34\cdot 10^{-4}\ M_{\odot}^{-1}$, respectively. Furthermore \citet{Mao11b} find that the delay time distribution that roughly follows a $t^{-1}$ power-law shape. Neither the integrated rate from the standard model nor from the variability models is consistent with these observations. 
Recent measurements in volumetric surveys however have shown lower rates; $ (4.4\pm0.2)-(5.0\pm0.2)\cdot 10^{-4}\ M_{\odot}^{-1}$  by \citet{Per12}, $(13\pm 1.5) \cdot 10^{-4}\ M_{\odot}^{-1}$ by \citet{Mao12}, and $(4-12)\cdot 10^{-4}\ M_{\odot}^{-1}$ by \citet{Gra13}.
It is unclear if the different observed integrated rates are due to systematic effects (for example overestimation of the cosmic star formation history or over-correction of dust extinction) or if there is a real enhancement of SNeIa in cluster galaxies \citep[see also][]{Mao12}. 
With these recent observations of the integrated rate, the long-standing problem of BPS studies predicting too low SNIa rates has reduced. 
The SNIa rates of our most optimistic models of low duty cycles are comparable with the lowest observed integrated rates, but the corresponding synthetic DTD shows a stronger decline with time than the observed DTDs.

The increase in the SNIa rate in the mass-transfer variability models compared to the standard model is limited by the formation of close binaries with low mass companions. This depends on our understanding of binary evolution. 
A comprehensive comparison of four BPS codes \citep[including SeBa, see][]{Too14} showed that differences between the predictions of BPS codes for low- and intermediate-mass stars are not caused by numerical effects in the codes, but by different assumptions for phases in stellar and binary evolution that are not understood well. When these assumptions are equalized, the synthetic populations of the four BPS codes are similar. Important assumptions (or uncertain processes) for the SD channel are the retention efficiency for WD accretion and CE-evolution \citep{Bou13, Too13, Too14, Cla14}. 
\citet{Bou13} shows that the effect of different retention efficiencies can effect the SNIa rate by a factor 3-4 to even more than a factor 100, which explains for a large degree the large disagreement in the predictions of the SD SNIa rate by different BPS studies. 
Regarding the poorly understood common-envelope phase, we have shown that mass-transfer variability can effect the SNIa rate to a comparable degree as CE-evolution \citep[][this paper]{Rui09, Men10, Bou13, Cla14}. Especially now that the gap between observed and synthetic SNIa rates has decreased, it is important to take uncertainties in binary evolution such as retention efficiency, CE-evolution and mass-transfer variability into account. 

\section{Conclusions}
We have studied the effect of mass-transfer variability on accreting WDs in binary companion stars. 
Long-term mass-transfer variability can be induced by e.g. irradiation of the donor star by the accreting WD or by cyclic variations of the Roche lobe from mass loss episodes \citep{Kni11}. The timescale of the variability should be longer than the thermal timescale of the non-degenerate surface layer of the WD so that the surface burning is affected. On the other hand, the timescale of the mass-transfer cycles should not be too long, such that the binary is not affected in any observable way (e.g. strong bloating of donor stars by irradiation). Currently observations hardly constrain the theoretical models of mass-transfer variability \citep[e.g.][]{Bun04} and therefore we have constructed a number of models rather than studying the details of a particular mass-transfer variability model. 
We show that long-term mass-transfer variability can significantly affect the accretion process and retention efficiency of mass transfer towards WDs.

Mass-transfer variability and accompanying enhanced retention efficiencies is likely to impact the properties of accreting WD binaries. 
We find that irrespective of the specific shape of the mass-transfer variability, 
for all variability models the WDs can effectively grow down to average mass-transfer
rates a factor of $\beta$ lower than in the standard scenario without
variability. As an example, we study the evolution of SNIa progenitors from the single-degenerate channel. We find that if mass-transfer cycles take place, the parameter space of systems that become SNIa events is increased towards low mass donor stars. Furthermore we find that the integrated SNIa rate increases by a factor of about 2-2.5, which is comparable with the lower limit of the observed rates \citep[see][]{Mao11b, Per12, Mao12, Gra13}. Variability models in which the maximum mass-transfer rate is not limited affect the SNIa rate less. 
In conclusion, mass-transfer cycles potentially lead to a new formation channel of SNIa events that can significantly contribute to the SNIa rate. 

\section*{Acknowledgements}
We thank Gijs Nelemans and Christopher Tout for useful comments which helped us to improve the manuscript.
This work was supported by the Netherlands Research Council NWO (grant VIDI [\#016.093.305] and [\# 639.042.813]) and by the Netherlands Research School for Astronomy (NOVA). 

\bibliographystyle{mn2e}
\bibliography{general}

\label{lastpage}

\end{document}